# Investigating disaster response through social media data and the Susceptible-Infected-Recovered (SIR) model: A case study of 2020 Western U.S. wildfire season

Zihui Ma*[1], Lingyao Li[2], Libby Hemphill[2], Gregory B. Baecher[1], Yubai Yuan[3]

[1]Dept. of Civil and Environmental Engineering, University of Maryland, College Park, MD
[2]School of Information, University of Michigan, Ann Arbor, MI
[3]Dept.of Statistics, Pennsylvania State University, State College, PA

* Corresponding author: zma88@umd.edu

**Abstract:** Effective disaster response is critical for affected communities. Responders and decision-makers would benefit from reliable, timely measures of the issues impacting their communities during a disaster, and social media offers a potentially rich data source. Social media can reflect public concerns and demands during a disaster, offering valuable insights for decision-makers to understand evolving situations and optimize resource allocation. We used Bidirectional Encoder Representations from Transformers (BERT) topic modeling to cluster topics from Twitter data. Then, we conducted a temporal-spatial analysis to examine the distribution of these topics across different regions during the 2020 western U.S. wildfire season. Our results show that Twitter users mainly focused on three topics: "health impact," "damage," and "evacuation." We used the Susceptible-Infected-Recovered (SIR) theory to explore the magnitude and velocity of topic diffusion on Twitter. The results displayed a clear relationship between topic trends and wildfire propagation patterns. The estimated parameters obtained from the SIR model in selected cities revealed that residents exhibited a high level of several concerns during the wildfire. Our study details how the SIR model and topic modeling using social media data can provide decision-makers with a quantitative approach to measure disaster response and support their decision-making processes.

# 1. Introduction

Effective disaster response is critical for affected communities to withstand and recover from the impacts of disasters (Tsou et al., 2017). However, in a disaster environment, situations can change rapidly, and unexpected issues can arise (Li, Bensi, et al., 2021; Tint et al., 2015), making it crucial to have timely information to support decision-making. In particular, decision-makers need up-to-date information on the status of critical infrastructure, search and rescue efforts, and emerging issues (Asimakopoulou & Bessis, 2011), to address the concerns and needs from the affected communities.

Conventional approaches to investigating disaster response include interviews and questionnaires (Zade et al., 2018), remote sensing (Marlier et al., 2022), and sensor networks (AlAli & Alabady, 2022). However, traditional data collection processes such as interviews, surveys, and other on-the-ground inspection activities, can be time-consuming and therefore may not provide timely information (Reis-Filho & Giarrizzo, 2022). Remote sensing technologies can aid disaster response with high spatial resolution (Thangavel et al., 2023; Yuan et al., 2022), but they may not be appropriate for disaster-prone regions with severe weather conditions. Remote sensors also do not reflect specific concerns or sentiments of affected communities.

As an alternative, social media platforms such as Twitter or Facebook provide a real-time and widespread platform for sharing information during disasters. Social media data reflect attention from the people who respond to disasters. Recent studies have highlighted the potential of using social media data to support disaster response (Kankanamge et al., 2020; Li, Wang, et al., 2023) due to several advantages, including real-time updates, data quantities, and potentially high spatial coverage of users (Reuter & Kaufhold, 2018; Wu & Cui, 2018). In addition, social media can help to bridge the gap between decision-makers and the affected communities by enabling affected people to seek assistance and share information, while also enabling disaster stakeholders to understand the needs from affected communities and respond accordingly.

Advancements in natural language processing (NLP) models have benefited researchers to leverage social media data to support disaster response. One common method is to use sentiment analysis to gauge the emotional status of affected communities (Ragini et al., 2018; Sufi & Khalil, 2022). Other approaches involve using topic modeling to identify frequently discussed topics on social media (Shan et al., 2019; Yuan et al., 2021), which can inform decision-makers about pressing concerns from the affected population. While many studies have used NLP tools to identify important topics from social media posts, few have employed a quantitative approach to analyze topic diffusion over time and space. In addition, given that most social media users are from urbanized areas (Barberá & Rivero, 2015; Mislove et al., 2011), analysis of social media activities may not entirely reflect the responses in areas with fewer users. To address these challenges, the present study used Twitter data to conduct an analysis of the disaster response during the 2020 western U.S. wildfire season. Two research questions were proposed.

- **RQ1 (response magnitude)**: How did Twitter users respond to the wildfire disaster, and what were the most common types of responses?
- **RQ2 (spread velocity)**: How urgent were these responses, and how did they spread over time and across different affected areas?

To address RQ1, we used Bidirectional Encoder Representations from Transformers (BERT) topic modeling to cluster tweets into relevant topics with geo-information. To address RQ2, we employed the Susceptible-Infected-Recovered (SIR) model to explore the magnitude and velocity of topic diffusion on Twitter. Our study is innovative for disaster response by leveraging

SIR, a valuable model from public health for understanding temporal dynamics. This model facilitates a quantitative approach to measure disaster response and their dynamics on social media during disaster propagation. Taken together, this study offers important insights for decision-makers to investigate the disaster response efforts in the affected areas and provides them with a method to quantify the most critical responses over time and across different areas.

## 2. Literature Review

### 2.1. Conventional methods to study disaster response

Wildfires, characterized by their significant impact, demand prompt and effective crisis response. Conventional approaches, such as phone calls (Madey et al., 2007), sensor networks (Erdelj et al., 2017), on-the-ground inspections (Stone et al., 2018), and interviews (Downey et al., 2013), have provided significant insights into disaster response. However, these methods are resource-intensive, necessitating significant time and effort for pre-deployment. Consequently, they may not be able to deliver quick insights in the context of large-scale disasters.

Advanced technologies have emerged as prominent tools for assessing the impacts of natural disasters with remote sensing taking center stage in recent years (Shafapourtehrany et al., 2023). These techniques include high-resolution optical imagery (Gold et al., 2019), synthetic aperture radar (SAR) (Ghosh et al., 2021), and light detection and ranging (LIDAR) (Aoyagi et al., 2021; Baris et al., 2021), which have demonstrated effectiveness in measuring post-disaster damages, monitoring recovery and reconstruction progress, and evaluating infrastructure conditions (Fang et al., 2023; Thomas et al., 2023; Zhou et al., 2023). However, it is important to note that remote sensing technologies may not provide insights into individuals' responses and emotions. They could also encounter limitations when deployed in disaster-prone regions characterized by severe weather conditions.

## 2.2. Social media applications in disaster response

Recent studies have highlighted the potential of social media data in facilitating disaster response through real-time updates and the extraction of insights from public discourse during crises (Fan et al., 2020; Karami et al., 2020). This approach harnesses the concept of "citizen-as-the-sensor" (Goodchild & Glennon, 2010) and has the advantages of rapidity, large quantity, potentially spatial resolution, and near real-time information (Li, Bensi, et al., 2021; Li, Ma, et al., 2021; Reuter & Kaufhold, 2018; Wu & Cui, 2018). Previous studies have predominantly focused on sentiment analysis (Beigi et al., 2016; Ragini et al., 2018; Yuan et al., 2020) and text classification (Karimiziarani & Moradkhani, 2022; Li, Ma, et al., 2021, 2023; Xing et al., 2019) when analyzing social media data.

Sentiment analysis techniques are often used to extract positive or negative perspectives from social media texts, aiming to understand users' emotions during a crisis. For example, Mandel et al. (2012) utilized supervised classification to conduct sentiment analysis with Twitter users' demographic information during Hurricane Irene. Their study revealed variations in concern levels based on gender and location, which highlighted the importance of prioritizing vulnerable groups. In another study, Zhang et al. (2019) employed semi-supervised learning to classify emotions into multiple categories and observed a correlation between emotional expression on social media and users' information-sharing behavior. They also observed that negative information spread more rapidly during the initial and outbreak phases of the 2017 Hurricane Irma.

While sentiment analysis can offer valuable insights into the emotional state of affected communities, it may not adequately measure behavioral responses during disasters. Another challenge is the absence of a baseline when assessing the impact of disasters using sentiment analysis (Yuan et al., 2020). Specifically, it remains unclear how disasters can influence sentiments within a local area, as local residents might consistently express positive or negative sentiments regardless of experiencing an event.

Text classification is often used to understand public opinions and behaviors. Researchers have employed supervised machine learning approaches to classify social media messages into relevant categories. For instance, Xing et al. (2019) utilized Convolutional Neural Network (CNN) classifiers on their annotated data to classify responses in disaster management, supporting rescue operations and addressing victims' needs. Karimiziarani and Moradkhani (2022) categorized social media texts during Hurricane Ian into multiple categories, including damage, evacuation, injury, help, and sympathy. Additionally, Huang and Xiao (2015) annotated 10,000 tweets and divided social media data into 47 topic subcategories to train supervised classifiers. Their findings indicate geographic differences in situational awareness during a disaster event.

Nevertheless, supervised classification methods can require significant time and effort to annotate data. To address this limitation, researchers have also explored unsupervised methods that can automatically identify topics without the need for pre-defined categories. One widely used unsupervised model is K-Means clustering, which was employed by (Basnyat et al., 2017) to detect topics during the 2016 Ellicott City flash flood, including "help," "rescue," "damage," and "casualty." Another popular unsupervised approach is Latent Dirichlet Allocation (LDA), which was applied by (Zhou et al., 2023) to investigate public concerns and demands during Hurricane Laura. Similarly, (Moghadas et al., 2023) used LDA to explore the topics of public discussion on social media during different phases of the 2021 Germany flood.

With the advancements in NLP techniques, researchers have applied Biterm Topic Modelling (BTM) (Zhang & Cheng, 2021) and transformer-based BERT (Chen & Lim, 2021; Prasad et al., 2023) to identify topics with social media text. These advanced NLP techniques have demonstrated their superior performance in such tasks (Grootendorst, 2022). However, only a few studies have applied these advanced topic modeling techniques to capture the dynamics of disaster response. In particular, few studies have used BERT to explore wildfire response, which highlights a research gap that requires further investigation.

**2.3. Social media application for the evolution of public concerns**

Studies that attempt to measure the evolution of public concerns are closer to the current study. Prior works in this field have examined the relationship between temporal changes in topics based on text volume and different stages of disasters (Ahn et al., 2021; Mihunov et al., 2022; Resch et al., 2018). For example, Huang and Xiao (2015) categorized social media messages into various topics corresponding to different disaster stages and analyzed the trends of these topics over time. Their findings revealed that the highest volume of messages fell into the "impact" category and occurred prior to the dissipation of the hurricane, which can be attributed to the large number of individuals affected by the hurricane. Similarly, Wang et al. (2015) used topic modeling based on Weibo messages and investigated the timely distribution of different topics during the 2012 Beijing Rainstorm. They identified the topic of "weather" as the most important one before the rainstorm and the topic of "loss" after the rainstorm.

Another common approach is to study the evolution of topics among social networks (Fu et al., 2020; Gallo et al., 2017; Ma et al., 2021). For example, Fu et al. (2020) proposed a framework that utilizes a co-word community evolutionary network to discover the relationship between the topic evolution and the rainstorm stage. Ma et al. (2021) developed a dynamic social community networking method based on key users and visualized the structural features at each stage of a snowstorm.

However, current approaches face two challenges. First, users from social media platforms may disproportionately represent urbanized communities (Barberá & Rivero, 2015; Mislove et al., 2011), potentially biasing the perception of the overall impact of a disaster. Second, while these approaches offer valuable insights into the evolution of topics in disaster response based on text volume or social network, there is a lack of quantitative methods to facilitate meaningful comparisons across different regions for decision-makers.

To address the research gaps, this study leverages an epidemic model called SIR to study the topic diffusion. The SIR concept was first introduced by Ross and Hudson (1917) in the early twentieth century, and then the model was developed by Goffman and Newill (1964). Goffman and Newill (1964) introduced the concept of "intellectual epidemic," which highlighted that individuals can be receptive to certain infections while resistant to others. The SIR model's advantage of considering user engagement as a participation ratio rather than real numbers has opened up new avenues for studying the evolution of topics on social media. Once an individual is "infected" with an idea, they propagate it to others in their network.

Prior studies have applied the SIR model to investigate information diffusion online (Kumar & Sinha, 2021; Woo et al., 2011; Woo & Chen, 2012; Xiong et al., 2012). For example, Woo et al. (2011) adopted the SIR model to model topic diffusion in the Jihadi forum. Woo and Chen (2016) extended the SIR model to analyze the spread of new topics in web forums, informing the success or failure of diffusion in the early stages and designing political and marketing strategies to enhance or reduce the chances of diffusion. These studies highlight the applicability of the SIR model in capturing the dynamics of information diffusion and its impact on online communities. However, its application in the context of disasters with social media data has not been thoroughly explored.

# 3. Data and methods

Figure 1 provides an overview of the research framework applied in this study. We used data from Twitter, topic modeling, and an epidemiological model to identify wildfire responses and how they changed over time. The process started with collecting and archiving English tweets related to the 2020 U.S. western wildfire season, as discussed in Section 3.1. Then, we used the registration location and content-based location of each tweet to enable spatial analysis, as outlined in Section 3.2. Thereafter, we employed the BERTopic tool to cluster tweets into distinct topics, with the aim of identifying categories in disaster response, as described in Section 3.3. Last, we used the SIR model to quantify the dissemination patterns of different topics among affected communities, as described in Section 3.4.

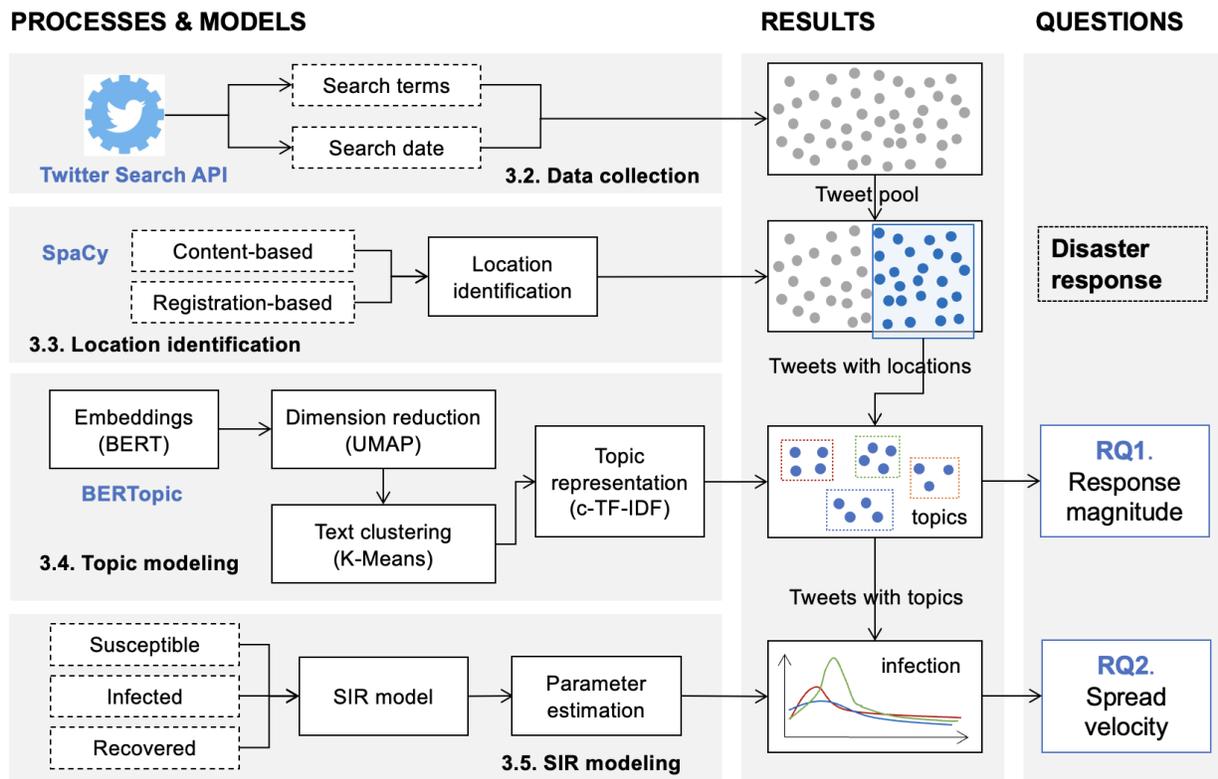

Figure 1. Graphical framework for the implementation of the current study.

## 3.1. Western Wildfire Season 2020

During the 2020 wildfire season, California, Oregon, and Washington were among the states hardest hit. In California, the wildfires consumed over 2 million acres as of September 2020, marking the largest wildfire season recorded in the state's modern history (Yan et al., 2020). In Oregon, the fires scorched over 1 million acres, leading to the evacuation of about 40,000 people and issuing evacuation warnings to an additional 500,000 individuals. At least 11 people lost their lives in the fires (Newburger, 2020; Schmidt & Friesen, 2020). In Washington, wildfires devastated over 713,000 acres, damaged 81 homes, and claimed one life. 2020 was recorded as having the highest number of individual fires compared to any other year (Decker, 2020; O'Sullivan, 2020).

## 3.2. Data collection

The 2020 U.S. western wildfire season generated widespread discussions on Twitter, presenting an opportunity to use Twitter data to study social media users' responses to the crises. We used the Twitter Search API with the keyword "wildfire" to collect relevant tweets between September 2 and October 4, 2020, while filtering for the U.S., where English is predominantly spoken. As no other significant wildfire events were reported during this period, we assumed that the collected tweets were related to the California and Oregon wildfires, although some fraction was possibly not. The resulting dataset comprised 932,325 records, including 175,946 original tweets.

## 3.2. Location identification

Location is critical for decision-makers investigating disaster response in specific areas. Twitter provides three location types: 1) content-based location, 2) posting location, and 3) registration location (i.e., home location) (Ao et al., 2014). Content-based location is the location mentioned in a tweet's text. For instance, if a tweet says, "*St. Helena Hospital in Napa County, California is* *being evacuated due to nearby wildfire*," it implies that the evacuation occurred at St. Helena Hospital, Napa County. Posting location, on the other hand, refers to the Global Positioning System (GPS) data that a user's device shares when publishing the tweet. Twitter enables users to disclose their home locations in their profiles; we call user-provided default locations "registration location."

We primarily used registration location to identify the location of a tweet. Zohar (2021) showed that the users' registration location could serve as potential proxies to infer event locations. Multiple prior papers have also used the registration locations of Twitter users to inform the locations of disaster events (Ozdikis et al., 2013; Sakaki et al., 2010). In addition, GPS location information was too rare to include in this sample (i.e., virtually no users shared their GPS locations). In our dataset, 117,221 tweets include registration location.

To supplement the location data, we also used content-based location, as a prior study showed that such data have good consistency with the locations of users (Paradkar et al., 2022). We used the SpaCy Named-Entity Recognition (NER) tool to extract content-based locations. NER is an information extraction sub-task that aims to identify and categorize named entities in unstructured text, such as names, organizations, and locations. This approach resulted in 41,502 tweets containing content-based location information. For tweets that contained both content and registration location, we kept only the registration location information. This left a total of 151,682 tweets with 36,963 original tweets for subsequent analysis.

### 3.3. Topic modeling

We employed a transformer-based topic modeling approach called BERTopic (Grootendorst, 2022) to analyze a large collection of tweets and identify common themes. Unlike traditional topic models (e.g., LDA) that rely on bag-of-words representations, BERT-based models can capture the semantic relationships and contextual meanings of words. Before applying the BERTopic model, we preprocessed the tweets using several NLP tools, including tokenization, sentence splitting, removal of stopwords, and removal of tags and special characters.

The BERTopic includes several steps to generate meaningful topic representations (Grootendorst, 2022). First, we used a pre-trained BERT model – the Sentence-BERT (SBERT) framework to convert each document into an embedding representation. The next step involved using a tool called Uniform Manifold Approximation and Projection (UMAP) to reduce the dimensionality of these embeddings. In this dimension reduction process, we considered two crucial parameters, namely "n_neighbors" and "min_dist," to reach a balance between local and global structure in the final projection. The "n_neighbors" parameter determines the number of neighbors whose local topology is preserved, with higher values preserving global distances. Meanwhile, the "min_dist" parameter influences the minimum distance between samples in the embedding, thereby affecting the spread of clusters. For our experiments, we used the default values of these parameters (n_neighbors =15, min_dist =0.01).

Additionally, to achieve reasonable visualization, we fine-tuned the "n_components" to 5. Moreover, we chose "cosine" as the distance metric to find the nearest neighbors during the process. After that, we used K-Means clustering to group the 36,963 original tweets into clusters. For our dataset, we chose 50 clusters as the optimal number, considering a balance between clustering granularity and topic interpretability. We tested $k = 20$, $k = 100$, and $k = 200$ as well. $k = 50$ gave us a reasonable balance between coherence and interpretability.

Finally, we used a custom class-based variation of Term Frequency-inverse Document Frequency (TF-IDF) (Eq. 1) to extract topic representations (i.e., representative words and tweets) from these clusters. This approach helps to generate coherent and meaningful topic representations.

$$w_{x,c} = \|tf_{x,c}\| * \log(1 + \frac{A}{f_x}) \qquad \text{Eq. 1}$$

in which,

$w_{x,c}$ = a term $x$ within class $c$

$tf_{x,c}$ = frequency of word $x$ in class $c$

$f_x$ = frequency of $x$ across all classes

$A$ = average number of words per class

We manually reviewed representative tweets and keywords within each topic and assigned the topic into a parent category. Two authors were involved in this process and continued refining the categories until reaching an agreement for each topic. Ours is not the first study to examine the topics of conversation during disasters. Once we had reviewed the topics BERT identified and labeled them manually, we compared our typology with prior works. Our goals were to verify our results and to ensure that we used common vocabulary to connect our work with existing literature on disaster response. The supporting literature was listed in Table A1 (see Appendix), and the final topics were detailed in Table 1. It should be noted that our study focused on disaster response; hence we did not analyze tweets from the following categories: "wildfire causes," "misinformation," "COVID-19 pandemic," and "other."

Table 1. Topic identification and examples of representative tweets and keywords.

| Cluster number | Representative keywords | Representative situation | Representative tweet example | Topic |
|---|---|---|---|---|
| 0, 1, 5, 9, 10, 12, 16, 21, 37 | smoke, health, lung, air, ash, sky, quality, unhealthy, orange, wind | Health risks, impacts on air quality | *These are not clouds. Wildfire smoke \*cough\** | Health impact |
| 4, 11, 15, 18, 20, 24, 27, 28, 29, 30, 35, 38, 48 | acre, burn, glassfire, death, kill, school, close, highway, missing, scorch | Infrastructure disruptions, injuries, fatalities, and other consequential social and economic damages | *Wildfire. right lane blocked in #Sherwood on 99w SB near SW Chapman Rd/SW Brookman Rd #PDXtraffic* | Damage |
| 31, 46, 47 | expect, latest, ready, prepare, devastation, real, map, live, time, announce | Live report, real-time map, and other tracking tools and methods | *California, Oregon, and Washington live wildfire maps are tracking the devastation in real time* | Monitoring |
| 2, 7, 22, 23, 25, 32, 43 | evacuate, firefighter, gov, rescue, level, order, battle, response, animal, governor | Evacuation plans, rescue efforts, and other response measures facilitating evacuation. | *All of Clackamas County is under Level 1, 2 or 3 evacuation order* | Evacuation |
| 8 | insurance, housing, risk, family, earthquake, disaster, safe, emergency, plan, homeowner | Infrastructure repair, emotion restoration, community rebuilding, and other recovery efforts | *RT @CAL_FIRE: Returning home after a wildfire can be difficult and the amount of damage is often unknown. Take the time to inspect the outside and inside of your home for any remaining dangers. Learn what to look for by visiting* | Recovery |
| 6 | relief, fund, donation, support, donate, effort, evacuee, victim, recovery, community | Donations, relief fund, and other humanitarian endeavors | *Please give if you can. Wildfire Relief Fund* | Humanitarian aid |
| 3, 13, 36 | party, pyrotechnic, spark, cause, climate, change, forest, management, tree, fuel | User report information about the causes of wildfires | *A couple's plan to reveal their baby's gender went up not in blue or pink smoke but in flames when the device they used sparked a wildfire that burned thousands of acres and forced people to flee from a city east of Los Angeles* | Wildfire causes* |
| 14, 26, 33, 42 | covid, pandemic, spread, virus, protest, resume, vaccine, live, arrested, china | User report information related to COVID-19 pandemic | *#Portland protests resume after wildfire hiatus with Ginsburg vigil, more vandalism* | COVID-19 pandemic* |
| 19 | antifa, arson, rumor, conspiracy, arrested, police, spread, blm, false, theory | User report information that related to the credibility of wildfire message | *Antifa setting wildfires is a conspiracy? See below links: ARSON-ANTIFA Conspiracy. Heh. Pt1* | Misinformation* |
| 17, 34, 39, 40, 41, 44, 45, 49 | fightwave, berry, madman, canadian, flight, foil, baked, photo, span, arrival | User report information that does not fit any above categories | *Oregon man arrested twice in 12-hour span for starting 'multiple' fires near Portland freeway* | Other* |

*Note:* *These topics did not reflect disaster response efforts and were not included in the following analysis.

### 3.4. Susceptible-Infected-Recovered (SIR) modeling

The Susceptible-Infected-Recovered (SIR) model addresses the outbreak and spread of diseases. The model specifies infection ($\beta$) and recovery ($\gamma$) rates and categorizes the population into three disease phases: (1) Susceptible, (2) Infective, and (3) Recovered (Harko et al., 2014). In the case of a disease outbreak, transmission begins when a few individuals become infected and subsequently come into contact with others. This leads to a continuous infection process until there are no more susceptibles or infectives left in the population. The SIR model has been applied beyond disease transmission to fields such as social networks (Kumar & Sinha, 2021), marketing (Woo & Chen, 2016), and informatics (Han et al., 2019), to help policymakers, marketers, and content creators better design communication strategies, predict the potential reach of information, and identify ways to optimize the spread of desired information or contain the dissemination of misinformation.

Topic diffusion on social media shares similarities with the process of disease transmission, as both rely on interpersonal contact for influence and spread (Bettencourt et al., 2005). Figure 2 illustrates how we adapted a general SIR model to study topic diffusion on Twitter. In the context of social media, the process of information distribution mirrors that of disease transmission. Woo et al. (2011) described this process in a study of extreme ideas in a web forum. Users start discussing topics that could potentially attract other users' attention. Users who show a certain level of interest in the topic become "susceptible" and therefore have the chance to contribute to the discussion by posting tweets. When users engage in the discussion, they become part of the "infected" subgroup. After a certain period, users lose interest or otherwise discontinue their participation in the discussions and join the "recovered" subgroup. Through the application of the SIR framework to model this dynamic of user engagement, decision-makers can gain insights into how information spreads and evolves during wildfire propagation, which further help optimize communication strategies, predict the potential reach of information, and devise methods to encourage desirable information dissemination.

The SIR process can be represented using three equations (Eq. 2, Eq. 3 and Eq. 4,), in which $S(t)$, $I(t)$, and $R(t)$ denote the size of the susceptible, infected, and recovered populations, respectively. $\beta$ and $\gamma$ represent the infection and recovery rates, respectively. The SIR model assumes that infection occurs when susceptibles come into contact with infectives, and the infection rate is dependent on the contact between the susceptible and currently infected users. In our study, it is worth noting that the model does not consider vital dynamics, such as birth and death rates, and therefore it assumes that the total population $N(t) = S(t) + I(t) + R(t)$ remains constant over time.

$$S(t) = \frac{dS}{dt} = -\frac{\beta IS}{N} \qquad \text{Eq. 2}$$

$$I(t) = \frac{dI}{dt} = \frac{\beta IS}{N} - \gamma I \qquad \text{Eq. 3}$$

$$R(t) = \frac{dR}{dt} = \gamma I \qquad \text{Eq. 4}$$

in which,

$S(t)$: the number of "susceptible" users involved in a topic at time $t$

$I(t)$: the number of "infected" users involved in a topic at time $t$

$R(t)$: the number of "recovered" users involved in a topic at time $t$

N: the total number of users involved in a discussion related to the crisis

$\beta$: infection rate

$\gamma$ : recovery rate

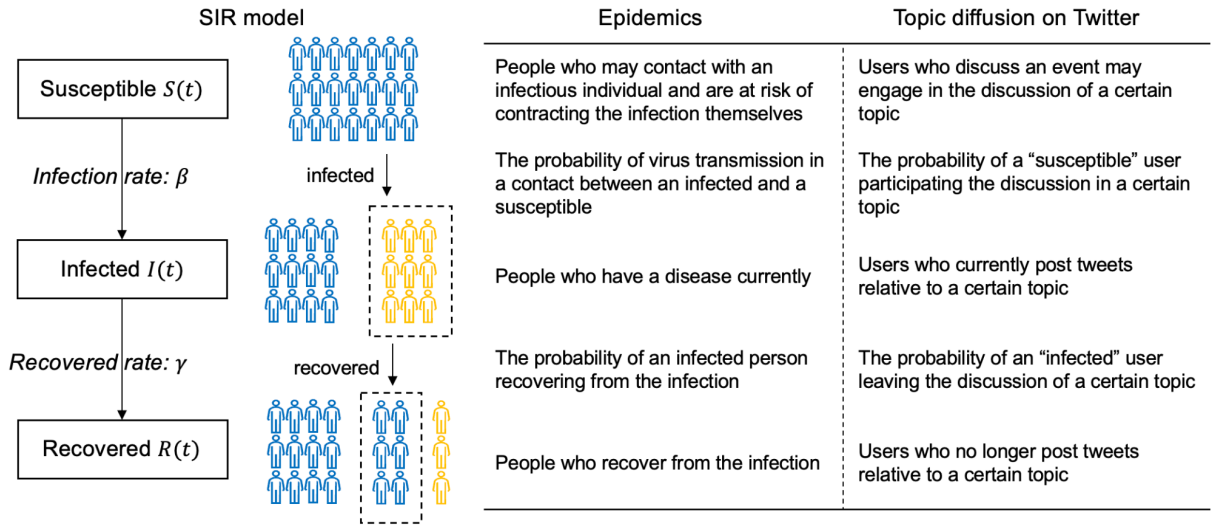

Figure 2. Application of SIR model to topic diffusion on Twitter.

After setting initial values of $S(0)$ and $I(0)$, the parameters $\beta$ and $\gamma$ were controlled to model the speed of infection and recovery. The number of infected users can be calculated as the product of $\beta$, $S(t)$, and $I(t)$. Along with the values of $\beta$ and $\gamma$, the system of ordinary differential equations governing the SIR model can be solved to obtain important metrics, such as the epidemic peak, which corresponds to the peak infected population of $I(t_*)$ at time $t_*$. The peak time $(t_*)$ represents the moment characterized by the highest level of user engagement and the most rapid diffusion of a particular topic. To get the estimation of parameters $\beta$ and $\gamma$, we applied the Least Square method to fit the dataset.

## 4. Results

Based on the proposed framework, we began by examining the overall changes in wildfire response across three heavily affected states: California, Oregon, and Washington. To showcase the variations in responses, we selected three representative cities from each state, namely Portland, Seattle, and Los Angeles, as elaborated in section 4.1. Additionally, we selected a specific wildfire to validate the applicability of the SIR model to certain locations, which allowed for a more granular insight into the feasibility of the proposed method, as presented in section 4.2. By analyzing Twitter activities surrounding different topics, we were able to show the progression of wildfires and gain valuable insights into the responses of affected communities at different stages of the wildfire.

**4.1. Analysis of the wildfire season**

4.1.1. Which topics received attention?

According to the six top-level topics defined in Section 3.3: "health impact," "damage," "evacuation," "monitoring," "humanitarian aid," and "recovery." The first three of these topics account for 88.37% of all tweets in our sample. The last, "recovery," garnered the least attention. Figure 3(a) illustrates the frequency of tweets in each topic over time and Figure 3(b) provides details regarding the distribution of tweets among topics. We included annotations of significant wildfire-related events to show that the topic diffusion and Twitter activities were associated with the timelines and locations of wildfires.

During the initial phase of the fire, there was limited public engagement evident on Twitter. However, after September 6, a notable surge in Twitter activity was observed, possibly due to the occurrence of several significant fires such as the August Complex fire and the Elkhorn fire in California (Chandler, 2020). The topic of "damage" gained significant prominence during the early days of September, the same time period as these fires. The pervasive smoke generated by the wildfires resulted in an eerie orange hue in the skies and affected regions, triggering extensive conversations surrounding the "health impact" of the wildfires.

Following a CalFire evacuation order issued on September 6, we observed multiple surges in tweet volume pertaining to the topic of "evacuation." These surges coincided with the implementation of large-scale evacuation orders in the affected areas (Hanna & Chavez, 2020; Schmidt & Friesen, 2020). One notable event occurred on September 9 when the city of Medford, CA issued an evacuation order due to the uncontrolled Almeda Drive fire (Davis, 2020). This order impacted over 80,000 residents, who were instructed to be on standby for immediate evacuation. On September 11, the Beachie Creek and Riverside fires merged in Oregon, affecting more than 10% of the state's population (Schmidt & Friesen, 2020). As a result, many local residents received evacuation orders. Around September 28, wildfires forced more than 70,000 residents of California state to evacuate (CNN, 2020), corresponding with another peak of tweet volume (Figure 3(a)).

In addition, we plotted the geographical topic distribution during this wildfire season, in comparison to the reported wildfires. Figure 4(a) illustrates the geographical locations of major wildfires sourced from the National Interagency Fire Center. Figure 4(b) showcases the distribution of topics, where the circle size corresponds to the tweet volume, and the color represents different topics.

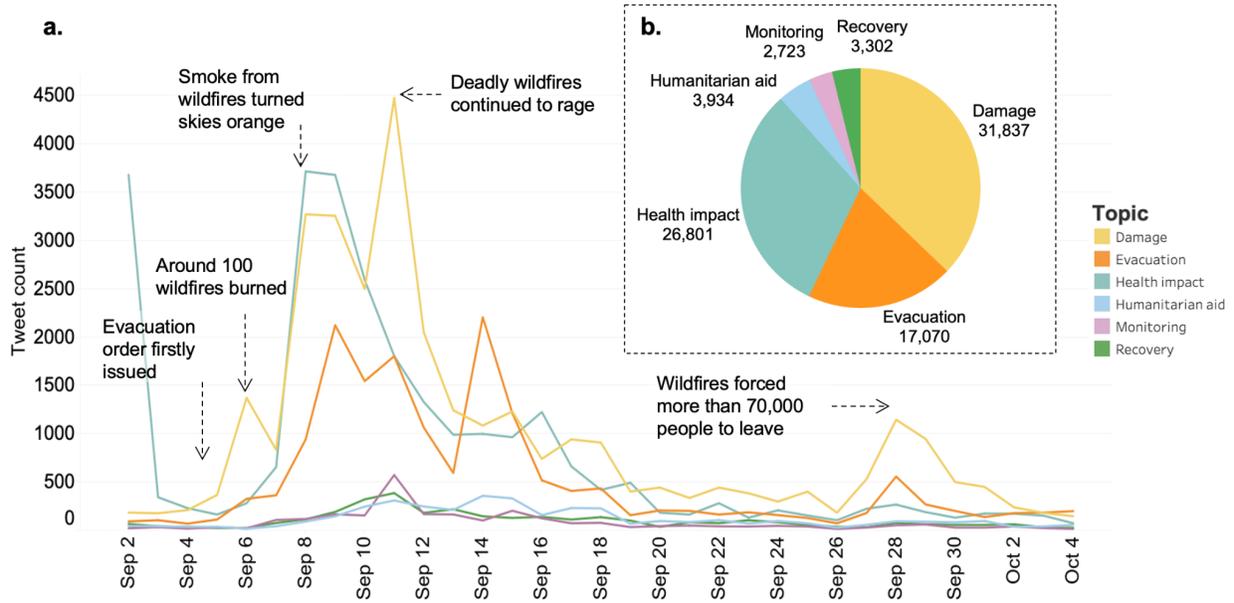

Figure 3 (a) Temporal dynamics of identified topics during the wildfire season. (b) Distribution of tweet volume among identified topics.

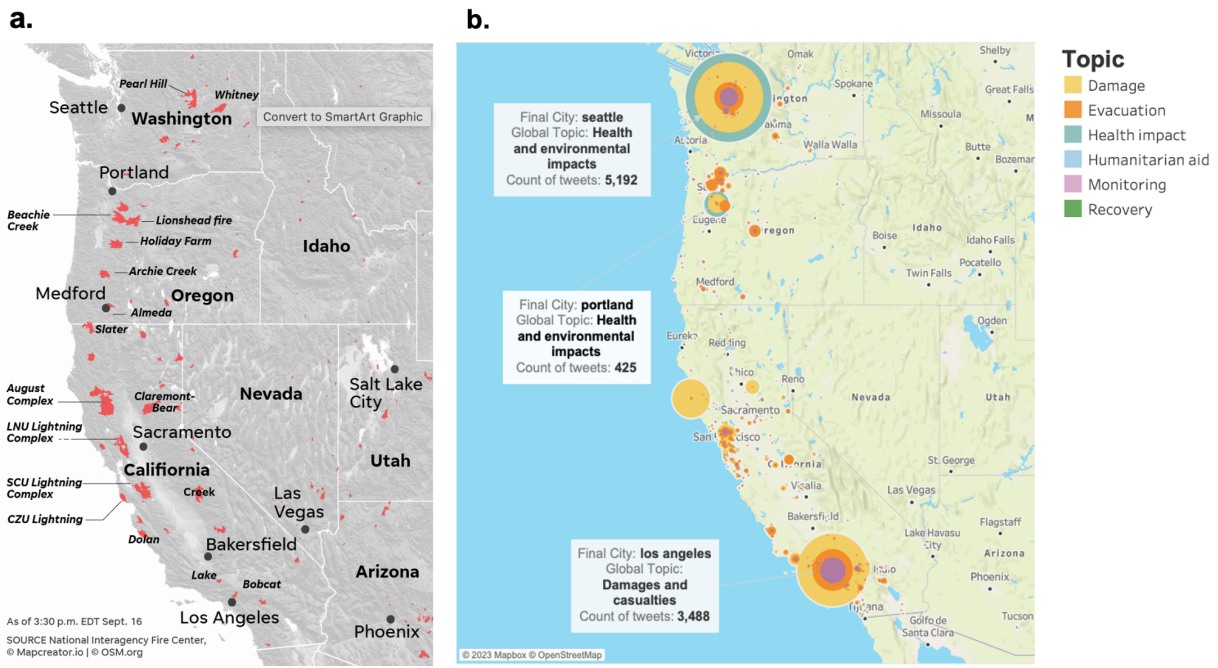

Figure 4. The map of (a) major wildfires in CA, WA, and OR (Gelles & Miller, 2020), and (b) topic spatial distributions during the wildfire season.

To further investigate the responses, we chose three representative cities from each state: Portland, OR; Seattle, WA; and Los Angeles, CA. We observed an alignment between Twitter activities and the fire locations, as depicted by the comparison between Figure 4(a) and Figure 4(b). This connection suggests that discussions on Twitter generally centered around the areas directly impacted by the wildfires. For example, in the initial week of September, California faced the devastating impact of several massive wildfires, which consumed approximately two million acres (NPR, 2020). Among those, the Bobcat Fire, located near Los Angeles, CA, resulted in substantial destruction and stood out as one of the largest wildfires ever documented in Los Angeles County. During that period, the map (Figure 4(b)) depicts discourse centered around Southern California.

Some topics, such as "health impact," "damage," and "evacuation," received similar levels of attention across regions. However, we also observed variations between locations concerning these topics. For example, the topic of "health impact" displayed a much larger volume in Seattle (5,192 tweets) than in Portland (425 tweets). This discrepancy suggests that Twitter users in Seattle were more actively engaged and displayed a greater awareness of health and environmental concerns associated with the wildfires compared to their counterparts in Portland. Moreover, the topic of "damage" exhibited a more widely distributed pattern across multiple locations on the map. This dispersion suggests that concerns regarding the extent of destruction and the impact on human lives were not limited to specific areas but were widespread throughout the affected regions.

4.1.2. Which topics spread quickly?

In this section, we conducted a computational experiment utilizing the SIR model to investigate user engagement across various topics from a city-level perspective. This simulation allowed us to analyze the magnitude and speed of topic diffusion. For the SIR analysis, each city's estimation consists of 32 daily observations. Table 2 summarizes the parameter estimation obtained from the SIR model. To interpret the results in Table 2, for example, the optimal parameter $\beta$ and $\gamma$ for the topic of "health impact" in Portland were estimated to be 2.65 and 2.04. This implies that the number of Twitter users who engaged in this topic increased by approximately 265 per day, and the number of users who left the discussion was 204 per day for every 100 "susceptible" users. The estimated peak time (12.25 days) suggested that this topic took around 12 days to become an influential topic.

Figure 5 illustrates the "infected" group in the SIR analysis for the selected topics, with the x-axis representing the date and time and the y-axis representing the participation ratio. We focused on the "infected" group as it helps understand the public response and their most pressing concerns during the crisis. Therefore, we excluded the other two simulated groups in Figure 5. The infection rate shows the speed of topic diffusion, while the participation ratio denotes the fraction of the "infected" users (i.e., who engaged in a topic) to the total Twitter users.

The estimates for the total user's engagement ($N$) in Portland appears approximately two times greater than that in Seattle and Los Angeles. Nevertheless, all six topics from these three selected cities showed an infection rate greater than one (Table 2), indicating that, on average, each "infected" user infected more than one "susceptible" user, resulting in an increase in the number of "infective" users on Twitter. This finding demonstrates that nearly all topics showed an outbreak (i.e., infection rate > 1) at the beginning of the study period.

Table 2. Parameter estimation using the SIR model.

| Location | Topic | Infection rate | Recovery rate | Peak time | Peak population |
|---|---|---|---|---|---|
| Portland (N=12,423) | Health impact | 2.65 | 2.04 | 12.25 | 348 |
| | Evacuation | 2.78 | 2.02 | 8.35 | 527 |
| | Damage | 3.18 | 2.53 | 9.8 | 286 |
| | Humanitarian aid | 4.01 | 3.59 | 11.75 | 72 |
| | Recovery | 12.68 | 10.84 | 8.3 | 137 |
| | Monitoring | 8.58 | 8.00 | 8.2 | 31 |
| Seattle (N=6,539) | Health impact | 2.15 | 1.23 | 6.4 | 719 |
| | Evacuation | 3.88 | 3.42 | 7 | 52 |
| | Damage | 2.37 | 1.51 | 8.95 | 497 |
| | Humanitarian aid | 4.23 | 4.07 | 9.5 | 7 |
| | Recovery | 9.09 | 8.30 | 8.3 | 33 |
| | Monitoring | 5.37 | 4.94 | 10.16 | 23 |
| Los Angeles (N=6,543) | Health impact | 3.36 | 2.69 | 8.56 | 141 |
| | Evacuation | 2.18 | 1.91 | 12.25 | 61 |
| | Damage | 1.16 | 0.91 | 15 | 181 |
| | Humanitarian aid | 5.67 | 5.24 | 10 | 20 |
| | Recovery | 3.53 | 3.16 | 10.55 | 39 |
| | Monitoring | 3.98 | 3.66 | 10.26 | 24 |
| Salem (N=436) | Health impacts | 4.22 | 3.39 | 4.2 | 10 |
| | Evacuation | 2.36 | 1.45 | 5.1 | 38 |
| | Damage | 2.16 | 1.21 | 5.7 | 51 |
| | Humanitarian aid | 1.46 | 1.28 | 12.35 | 7 |
| | Recovery | 0.00 | 0.07 | NA | NA |
| | Monitoring | 0.00 | 0.08 | NA | NA |

Based on Figure 5, Twitter users focused on three topics in wildfire response, evidenced by their higher participation ratios, which include "health impact," "evacuation," and "damage." However, the patterns of topic diffusion across the three areas were different. For example, users from Seattle expressed more concern about "health impact" than the other two cities, illustrated by a larger peak participation ratio (Figure 5(a)). For the topic of "evacuation" (Figure 5(b)), Portland was estimated to have the highest peak population with 527 with the highest participation ratio, while Los Angeles and Seattle had smaller peak populations with 61 and 52, respectively. For the topic of "damage" (Figure 5(c)), while Portland shows the highest infection rate ($\beta = 3.18$), its estimate of peak population was not as high as the other two cities. Moreover, one suspect was that the higher proportion of engaged users from Seattle might be associated with more severe damages inflicted to that area.

We further observed the other three topics, including "humanitarian aid,", "recovery," and "monitoring," had a comparatively lower level of user engagement, as evidenced by a smaller participation ratio in Figure 5. Among the three selected cities, Portland displayed the highest infection rate and recovery rate (Table 2) under the topic of "recovery" (Figure 5(e)). We also observed that the topic of "recovery" in Seattle (red line) and Portland (blue line) did not emerge at the beginning of our study period but instead appeared on day 5.

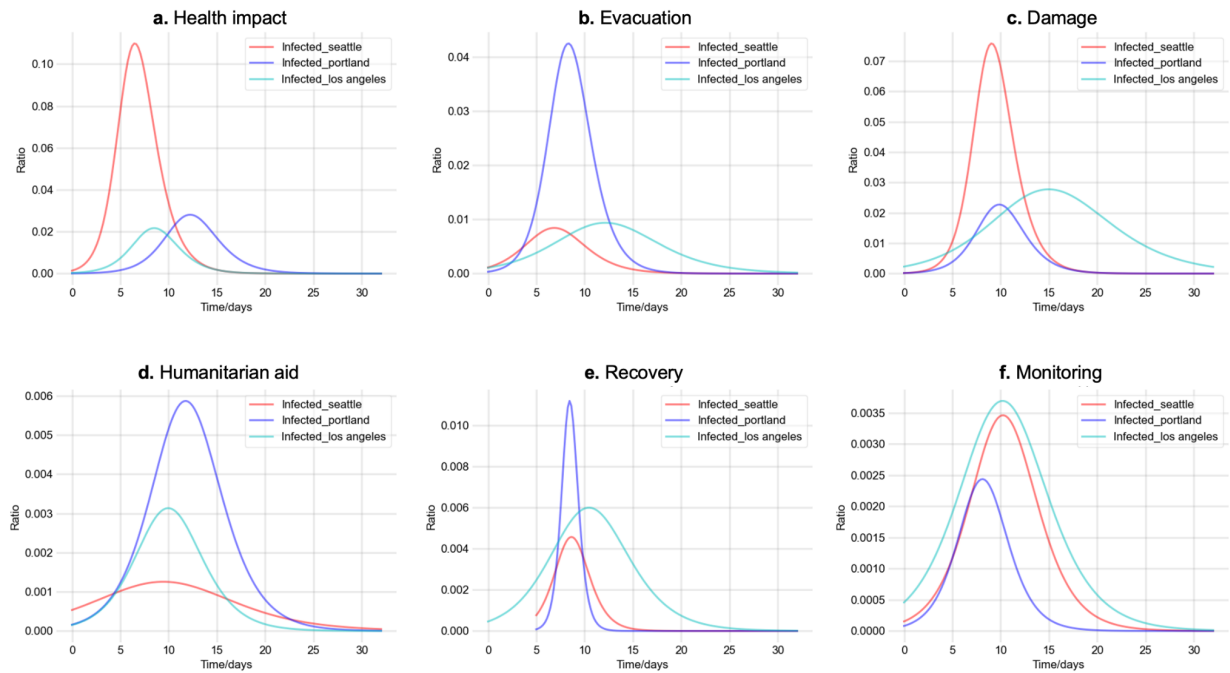

Figure 5

Figure 5. Infection curve in the SIR model for Portland, Seattle, and Los Angeles under the topic of (a) health impact, (b) evacuation, (c) damage, (d) humanitarian aid, (e) recovery, and (f) monitoring (day 0 corresponds to September 2, 2020).

## 4.2. The Santiam fire

Considering the varying policies and strategies adopted by different state emergency offices to manage wildfires, this section aims to investigate response patterns within a specific event on a localized scale. To achieve this, we chose the Santiam fire as a case study. The Santiam fire burned in northwest Oregon and caused significant damage to local communities (Bay Area News Group, 2020). The fire was labeled as "Santiam fire" when it combined three nearby fires that occurred early on September 8, 2020, and was not fully contained until the end of 2020. However, the most severe days of the wildfire lasted until September 10 in Portland area (Urness & Radnovich, 2020). Therefore, we focused on a four-day timeframe spanning from September 7 to 10 to examine its impact on the discussions within local communities.

4.2.1. Which topics received attention?

Figure 6 illustrates the Twitter activity for each topic during this period. Based on Figure 6, we observed an association between the trends of Twitter activity and the stages of the Santiam fire. For example, we observed an increase in Twitter volume on September 8, 9, and 10. Topics that dominated the discussion include "health impact," "damage," and "evacuation," which appeared consistent with our previous discussion in section 4.1.

On September 7, Twitter discussions centered on the topic of "health impact" in Portland (tag A in Figure 6(a)), which coincided with the initiation of the Santiam fire. However, there was limited discussion in Salem (tag B in Figure 6(a)) as there were rarely tweets presented on the map. On September 8, the Santiam fire merged with three other nearby fires, including Beachie Creek fire, Lionshead fire, and P-515 fire (Bay Area News Group, 2020). Meanwhile, tweet volume saw a dramatic increase in response to this wildfire, as evidenced by the sudden appearance of circles on the map. Moreover, as the state declared an emergency on the same day (Miller, 2020), the topic of "evacuation" gained significant discussion on Twitter, as depicted in Figure 6(b).

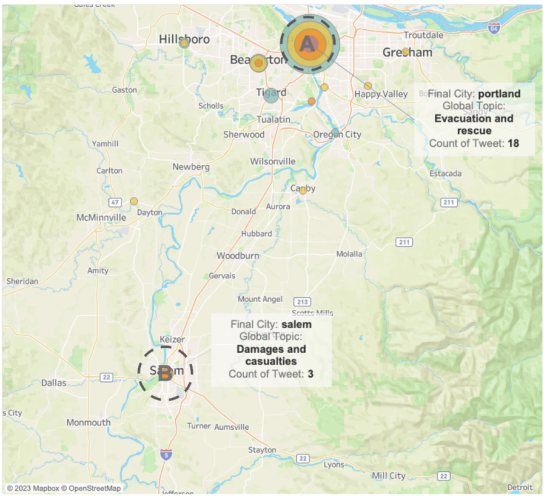
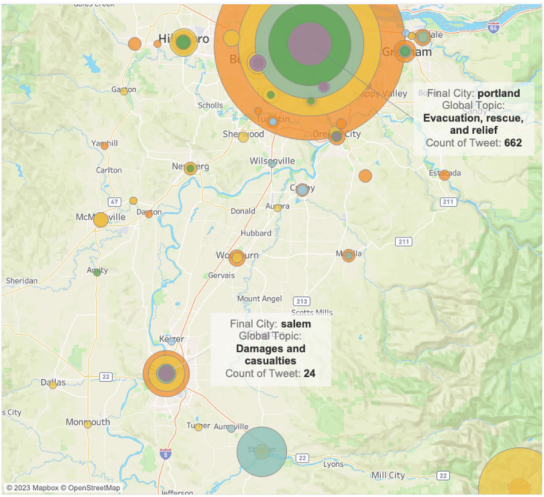
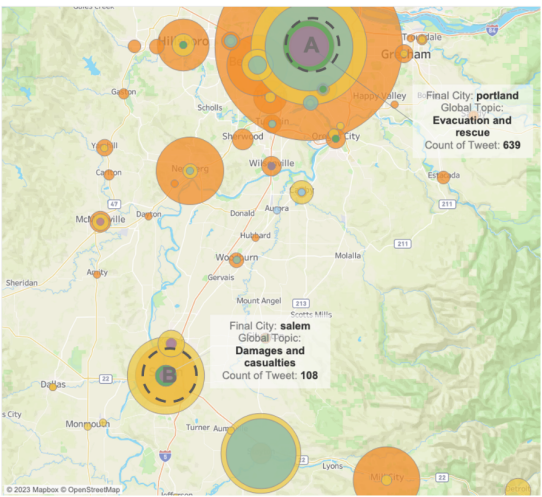
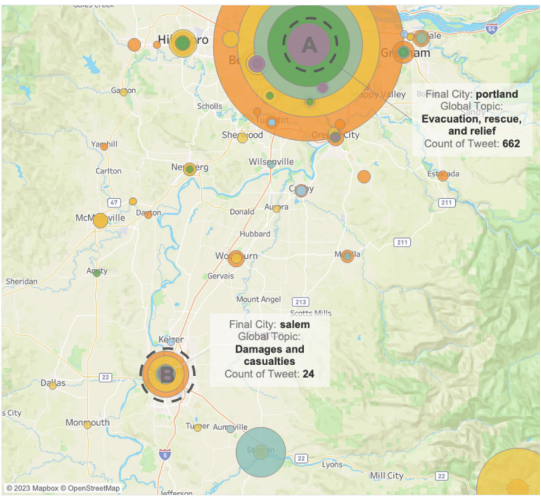

Figure 6. Integrated topic distributions for Santiam Fire on (a) September 7, 2020, (b) September 8, 2020, (c) September 9, 2020, and (d) September 10, 2020.

Such discussion continued to grow on September 9 in both the Portland and Salem areas, as the Santiam Fire spread westward, causing extensive damage across several cities in Marion County (around tag B in Figure 6(c)). Correspondingly, there was a significant surge in the topic of "damage," as indicated by the yellow circles scattered throughout the map. However, Salem (tag B in Figure 6(c)), situated at the center of the Santiam fire, showed comparatively less attention on the topic of "evacuation" on Twitter possibly due to a smaller number of affected residents residing in this area.

On September 10, tweet volume showed a (Figure 6(d)) slight decrease. At this stage, residents in Salem (tag B in Figure 6(d)) continued to discuss the topic of "damage." Furthermore, we observed an expansion of the "recovery" topic in Portland (tag A in Figure 6(d)), suggesting that the community in this area might have started to cope with the aftermath of the wildfire and paid attention to fire recovery efforts.

4.2.2. Which topics spread quickly?

Despite the rapid growth of the Santiam fire into a large blaze from September 7 to September 10, its impacts extended throughout the entire wildfire season. As a result, our SIR analysis encompassed the entire daily dataset but primarily focused on the two affected cities, Portland and Salem, to validate the application of SIR model. Their parameter estimation is listed in Table 2, and the simulated SIR result is illustrated in Figure 7. As mentioned in the previous section, we only analyzed the dynamics of infection groups for each topic.

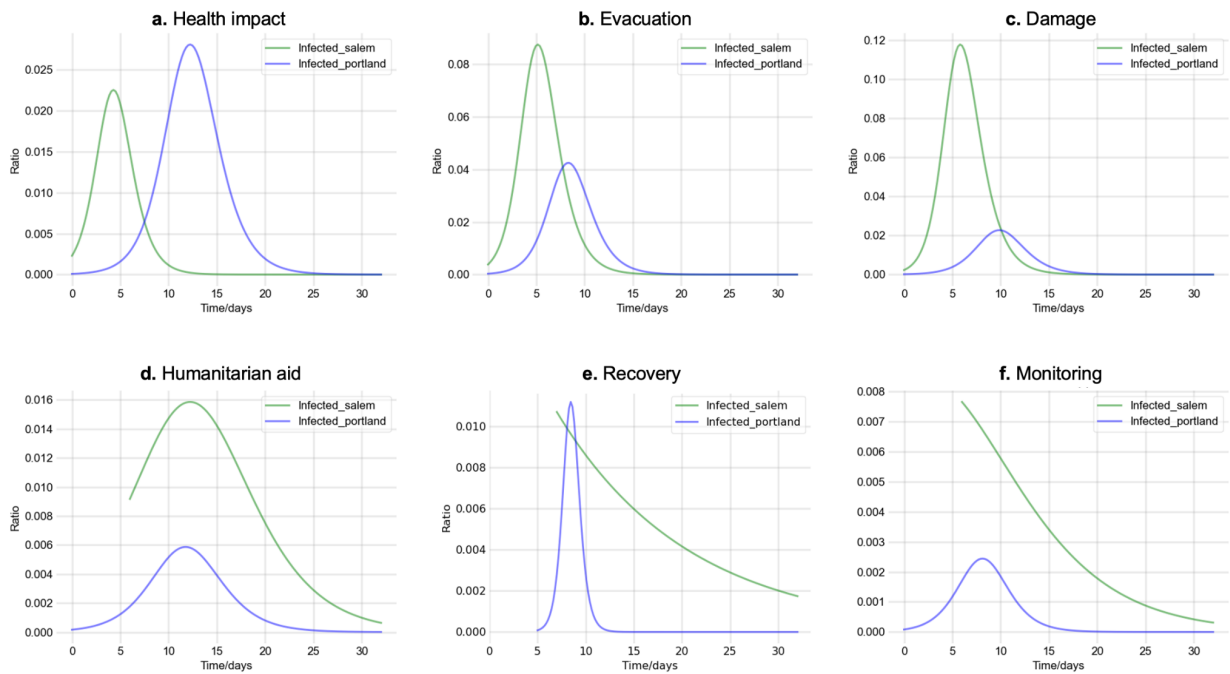

Figure 7. Infection curve in the SIR model for Portland and Salem under the topic of (a) health impact, (b) evacuation, (c) damage, (d) humanitarian aid, (e) recovery, and (f) monitoring (day 0 corresponds to September 2, 2020).

While the number of Twitter users in Salem is significantly smaller than Portland, the infection rates from these topics are not consistently lower. For example, as Figure 7 shows, the topic of "health impact" shows a greater infection rate in Salem and peaked at around 4.2 days. This observation highlights the capability of our SIR model to capture the level of responses from a specific location, even in cases where the number of social media users is relatively small. We also observed that users from both areas focused on three topics in wildfire response, including "health impact," "evacuation," and "damage." In particular, Salem showed the highest concern over the topic of "damage" (Figure 7(c)) with a peak population of around 51 (participation ratio ≈ 0.12).

However, it is important to acknowledge that a higher infection rate does not necessarily correspond to a higher participation ratio. The infection rate solely indicates the speed at which a topic spreads at a specific time point. For instance, in the case of the topic "health impact" (Figure 7(a)), Portland showcases a lower infection rate (Table 2) but a higher participation ratio in comparison to Salem. This indicates that a larger proportion of users in Portland actively engaged in this topic – it had broader participation but did not spread. Similarly, when considering the topic of "evacuation" (Figure 7(b)), Salem displays a higher participation ratio but a lower infection rate (Table 2) compared to Portland.

Among the three remaining topics, the topic of "humanitarian aid" in Salem (Figure 7(d)) began trending on Twitter around the sixth day (September 7), with a significant increase in user engagement that reached its peak at approximately 12.35 days (around September 13). Moreover, we observed a decline in dissemination for the other two topics, namely "recovery" (Figure 7e)) and "monitoring" (Figure 7(f)) in Salem. This is evident from their infection rates being equal to 0 and recovery rates being greater than 0. Consequently, the specific peak time and peak population for these two topics remain unknown in this case.

## 5. Discussion and implications

This study exemplifies the feasibility of employing the BERTopic and SIR models to comprehensively capture public responses during a wildfire season, both the magnitude of and velocity of topic diffusion. The results also highlight the potential of leveraging social media data to provide near-real-time information regarding public concerns and responses across different affected locations. The following key findings emphasizes the significance of our research.

Our findings first show that Twitter users expressed significant concerns regarding three topics during the wildfire season. These were: "health impact," "damage," and "evacuation." The convergence of these topics reflects the severity of the fires. Additionally, the wildfires inflicted extensive and severe damages, leaving a profound impact on both the environment and public health. This includes detrimental effects on air quality, as well as the destruction of crucial infrastructure.

The results suggest that the speed at which a topic spread in a particular area could reflect public responses to the wildfire crisis. The faster the speed, the greater the concern. Residents across all three states were aware of the risks associated with "health impact" and "damage." With regards to Santiam fire, the variation in topic diffusion was consistent with the wildfire propagation, while it was linked to people's awareness disparities.

The results also suggest that communities could exhibit varying levels of responses in the face of wildfires. For example, individuals in Seattle demonstrated a heightened concern regarding health impact, whereas those in Portland demonstrated a greater attention to damages, injuries, and fatalities. This divergence underscores the distinct impacts experienced by different communities and may reflect varying levels of awareness and understanding regarding the calamity.

Finally, the temporal-spatial analysis suggests that topic diffusion aligns closely with the locations and timeline of the fires. The level of user engagement within each topic serves as a reflection of how wildfires impacted the local areas. For instance, the topic of "evacuation" appeared in closer proximity to the Portland area as the Santiam wildfire largely affected local communities. Additionally, the examination of Twitter activities uncovered that the diffusion of topics corresponded to distinct stages of wildfires, as depicted in Figure 6. This temporal pattern signifies that the discussions and concerns expressed on Twitter dynamically mirrored the progression of the wildfires, highlighting their real-time nature and responsiveness to unfolding events.

**5.1. Implications for wildfire disaster response**

This study presents multiple implications for decision-makers to consider when addressing wildfire response. These implications stem from the concerns and perspectives shared by affected individuals, as conveyed through social media platforms. By taking these implications into account, decision-makers can gain valuable insights into the needs and concerns of the affected population.

First, decision-makers can leverage the topic-based SIR model as a quantitative approach to measure the magnitude and velocity of topic diffusion. This proposed approach offers two advantages compared to conventional methods that rely only on tweet volume. The topic-based SIR model captures the temporal trends and peak periods of these concerns, thereby providing decision-makers with insights to identify the most critical concerns within the affected communities. Additionally, it enables a quantitative measurement of the impact across different geographical areas, irrespective of variations in the number of users engaging in the discussions. This is important as there are fewer social media users in less urbanized areas (Barberá & Rivero, 2015; Mislove et al., 2011).

Second, decision-makers can leverage social media data to understand the dynamics of information evolution during wildfire seasons. By doing so, they can adapt their communication strategies to ensure timely and effective dissemination of information, which can help promote public safety and foster the development of resilient communities. For example, our topic-based SIR model shows that people in different areas showed varying levels of concerns regarding different topics. This information can help decision-makers to tailor strategies to meet the critical needs of the affected communities.

Third, decision-makers should address those concerns with high infection rate or participation ratio. This aspect holds particular significance as it empowers decision-makers to effectively allocate resources to areas that have been heavily affected. To illustrate, our findings show that users from Washington and Oregon states were attuned to the health impact, while users from California were more concerned with the extent of afflicted damage and loss. Therefore, it is crucial for wildfire decision-makers in Washington and Oregon to provide timely information that addresses wildfire risks, thereby improving public awareness and preparedness for the health-associated impacts prevalent in those regions.

Last, we suggest decision-makers taking notes of topics with low infection or participation ratio when dealing with disaster response. Despite the lack of significant attention on social media platforms, such topics can still play a critical role in wildfire disaster response. For example, in all three states, there was minimal concern regarding the monitoring and updates of wildfires. However, enhancing wildfire monitoring is of utmost importance as it enables timely notifications to the affected communities about the current wildfire status. This, in turn, facilitates effective disaster preparedness and evacuation efforts.

**5.2. Limitations and future work**

Several limitations need to be highlighted. One limitation arises from the identification of locations. In this study, we relied on registration location and content-based location to demonstrate the disaster response. There remains a possibility that users may post tweets implying situations unrelated to their registration location. Furthermore, content-based location identification relies on the NER model, which could result in misidentification of locations. One area for future research could involve the application of more advanced tools, such as large language models (LLMs), to enhance the identification of locations. Additionally, it would be valuable to conduct an assessment of introduced biases or errors when utilizing these location identification methods.

The second limitation pertains to the topic modeling process employed in the study. The BERTopic algorithm clusters tweets based on their textual similarity, but it is important to acknowledge that it could potentially group tweets with similar words but different contextual meanings into the same cluster. Furthermore, we observed instances where certain cluster topics encompassed tweets related to distinct topics. This occurrence was particularly prevalent in clusters with a substantial number of tweets. To mitigate this limitation, future work could consider exploring other LLMs such as integrating GPT-based models (Chen et al., 2021) with the topic modeling.

Moreover, it is essential to recognize that manual interpretation of topics can introduce uncertainties into the final topic classification. Due to diverse backgrounds, individuals may possess varying conceptual understandings of topic descriptions. In many instances, researchers assign high-level descriptors to each topic, with the topic model encompassing finer distinctions than existing classification schemes (Baumer et al., 2017). Indeed, the investigation of the relationship between interpretive approaches and computational analysis techniques could deserve future research endeavors.

Another limitation is from the SIR model. The traditional SIR model treats information channels or topics independently. In real scenarios, multiple topics could diffuse and interact simultaneously, but such interaction effects were not addressed within the traditional SIR model (Razaque et al., 2022; Wang et al., 2021). Moreover, the traditional SIR model assumes a homogeneous mixing of the population, which may not accurately represent our application. For example, it does not consider other statuses of users, such as "returning" users who leave and later rejoin the topic. Additionally, Twitter data may exhibit multiple peaks, whereas the current model only accounts for one peak in the process.

It is also important to acknowledge that relying on postings on Twitter could introduce inherent bias. The individuals who wrote tweets might not necessarily represent the broader public response during wildfire disasters. For instance, research has consistently demonstrated that young, educated, and urbanized individuals tend to be more active in posting comments on social media platforms due to their familiarity with these technologies (Barberá & Rivero, 2015; Mislove et al., 2011). Furthermore, some individuals might choose alternative social media platforms like Facebook (Maas, 2019) or TikTok (Basch et al., 2022) and use videos or images to share their observations or experiences during disasters. These factors can potentially influence the quality of data preparation and introduce biases into the results. Therefore, exploring data from other social media platforms, such as Facebook or TikTok, or incorporating data from more representative sources like sample surveys, could be a valuable avenue for future research.

There are two additional avenues of future research that deserve investigation. First, we plan to use the SIR model to assess situational awareness during disasters and validate its estimation using community vulnerability score (FEMA, 2022b) or community resilient index from FEMA (FEMA, 2022a). This holds particular significance as it enables decision-makers to quantitatively measure situational awareness levels across communities and proactively address the associated risks. By leveraging the SIR model, decision-makers can also enhance their preparedness efforts and allocate resources to mitigate the impacts of disasters.

Second, we plan to examine the generalizability of the proposed model to other weather-related disasters, such as hurricanes and flooding. By testing the applicability of our model beyond wildfire scenarios, we can assess its effectiveness in investigating public response and recovery dynamics in different disaster contexts. This exploration could broaden the understanding of the proposed model's utility and shed light on the potential of using social media data to assess disaster response and situational awareness.

## 6. Conclusions

Social media has the potential to support disaster response and recovery efforts. To demonstrate this potential, our study collected and analyzed Twitter postings during the 2020 western wildfire season. By employing BERT-based topic modeling and the SIR model, we developed a quantitative model to measure topic diffusion across different regions during this wildfire crisis. Through our topic analysis, we discovered that Twitter users concentrated on three topics, "health impact,"

"damage," and "evacuation." This observation highlights the key areas of concern for the public during wildfire disasters. Additionally, our analysis revealed a strong association between the patterns of public responses and the locations and timelines of the wildfires. The divergence observed across different areas not only underscores the distinct impacts experienced by various communities but also reflects their varying levels of awareness and understanding of the calamity.

The proposed model offers valuable insights into user engagement and topic diffusion, providing decision-makers with a quantitative method to measure public responses during a disaster across different regions, regardless of variations in the number of users engaged in the discussions. The findings further provide invaluable insights for decision-makers to identify and address the most critical risks and concerns within the affected communities, allowing for targeted and effective response strategies to be implemented.

## Declaration of competing interest

The authors declare that they have no known competing interests or personal relationships that could have appeared to influence the work reported in this paper.

# Appendix

Table A1. Supported literature with corresponding topics.

| Author and Year | Disaster type | Topic in disaster response |
|---|---|---|
| Zhou et al. (2023) | Hurricane | "Advisory," "Casualty," "Damage," "Relief," "Information source," "Emotion," "Animal" |
| Gründer-Fahrer et al. (2018) | Flood | "Information," "Help," "Alertness/relief," "Relaxation" |
| Ahn et al. (2021) | Earthquake | "Occurrence of earthquake," "Place of earthquake," "Impact of earthquake," "Breaking news with video materials," "Magnitude of earthquake," "Facts and perceptions of earthquake," "Support and preparedness" |
| Karami et al. (2020) | Flood | "Victims," "Damage and Costs," "Drinking Water," "Insurance," "Homelessness," "Road Damage," "Roof Damage," "Bridge Damage," "Flood Report," "Power Lost," "Animal" |
| Wang et al. (2015) | Rainstorm | "Traffic," "Weather," "Disaster Information," "Loss and Influence," "Rescue Information" |
| Fan et al. (2020) | Hurricane | "Infrastructure and utility damages," "Affected and injured individuals," "Rescue, volunteering, or donation effort," "Other relevant information" |
| Karimiziarani and Moradkhani (2022) | Hurricane | "Caution," "Damage," "Evacuation," "Injury," "Help," "Sympathy" |
| Jin and Spence (2021) | Hurricane | "Medical support," "Resilience," "Food/water supply," "Forecast/update," "Rescue," "Concerns," "Aid," "Fundraising," "Evacuation," "Relief" |
| Huang et al. (2022) | Typhoon | "Warning," "Damage," "Work and life," "Temperature," "Concern and fear", "Traffic," "Caution and advice," "Gratitude," "Weather" |